\documentclass[pra,showpacs,twocolumn,graphics,floatfix,mathbbm,
superscriptaddress,nofootinbib]{revtex4}
\usepackage{amsthm}
\usepackage{amsmath}
\usepackage{latexsym}
\usepackage{amsfonts}
\usepackage{amssymb}
\usepackage{xcolor}
\usepackage{bbm,dsfont}
\usepackage{graphicx}
\usepackage{mathrsfs}
\usepackage{mathbbol}
\usepackage{hyperref}
\usepackage{enumerate}
\usepackage{MnSymbol}
\usepackage{bbding}
\usepackage{float}
\usepackage{soul}
\usepackage{ulem}

\begin{document}

\title{Quantum  synchronization in dimer atomic lattices}

\author{Albert Cabot}
\affiliation{IFISC (UIB-CSIC), Instituto de F\'isica Interdisciplinar y 
Sistemas 
Complejos, Palma de Mallorca, Spain}

\author{Gian Luca Giorgi}
\affiliation{IFISC (UIB-CSIC), Instituto de F\'isica Interdisciplinar y 
Sistemas 
Complejos, Palma de Mallorca, Spain}

\author{Fernando Galve}
\affiliation{IFISC (UIB-CSIC), Instituto de F\'isica Interdisciplinar y 
Sistemas 
Complejos, Palma de Mallorca, Spain}
\affiliation{ I3M (UPV-CSIC) Institute for Instrumentation in Molecular Imaging, 
Universidad Polit\'ecnica de Valencia, 46022, Spain}

\author{Roberta Zambrini}
\affiliation{IFISC (UIB-CSIC), Instituto de F\'isica Interdisciplinar y 
Sistemas 
Complejos, Palma de Mallorca, Spain}

\begin{abstract}

Synchronization phenomena have been recently reported in the quantum realm at atomic level
due to collective dissipation. In this work we propose a dimer lattice of trapped atoms realizing a dissipative spin model where 
quantum synchronization occurs instead in presence of local dissipation. Atoms synchronization is enabled by
the inhomogeneity of staggered local losses in the lattice and is favored by an increase of spins detuning. 
A comprehensive approach to quantum synchronization based on different measures considered in the literature
allows to identify the main features of different synchronization regimes. 

\end{abstract}

\maketitle

Spontaneous synchronization (SS) among different interacting units 
is a paradigmatic collective phenomenon arising 
in a broad range of contexts \cite{ClassSync}. 
 In the last decade 
it has been explored into the quantum regime, which triggered novel questions related 
to the essence of this phenomenon and to its non-classical 
signatures. 
The very same definition of quantum synchronization has led to a variety of approaches 
and measures 
\cite{ZambriniRev,leHur,SyncCooling,SyncExperiment,SyncHO,GLG1,GLG2,Synchnetw,Solano1,manzanoSciRep,Bellomo,SyncIonsCorr,Mari,lee1,lee2,walter,SyncAtomicEnsemb,SyncDipoles,Maser,messina},
that can be broadly categorized as (i) time correlation of the dynamics of 
$local$ quantum observables, whose occurrence can be compared with 
quantum correlations \cite{leHur,GLG1,GLG2,SyncHO,Solano1,manzanoSciRep,Bellomo,SyncIonsCorr,Synchnetw}; 
or as (ii) reduction 
of noise in some collective variables, being then itself a form of $global$ quantum 
correlations \cite{Mari,lee1,lee2,walter,SyncAtomicEnsemb,SyncDipoles,Maser,messina,SyncCooling,SyncExperiment}.

The study of quantum SS has enriched the perspective on this 
phenomenon in different dynamical regimes. 
Classical SS has been broadly studied in 
self-sustained oscillators, encompassing regular periodic, but also chaotic 
 and stochastic evolutions \cite{ClassSync,Boccaletti,Arenas}. 
Quantum self-sustained oscillators can also display quantum SS, as 
reported in Van der Pol oscillators \cite{lee1,lee2,walter}, optomechanical systems \cite{Mari,ZambriniRev,Marquardt}, micromasers \cite{Maser}, spin-1 systems \cite{bruder}, and ions  
\cite{SyncIonsCorr}. 
Apart from this, different dynamical scenarios have been explored in the quantum regime, 
leading either to SS in the steady state or in transient dynamics, as
 in steady superradiant
emission \cite{holland2012,SyncCooling,SyncAtomicEnsemb,SyncDipoles,SyncExperiment}  and in presence of decoherence free subspaces \cite{Synchnetw,manzanoSciRep}, relaxing networks of harmonic oscillators  \cite{SyncHO,Synchnetw} or spins  \cite{leHur,GLG1,GLG2,Solano1,Bellomo,messina}. 
In atomic systems genuine quantum features of synchronization come into play,
as in superradiant lasers \cite{holland2012},
in supercooling \cite{SyncCooling}, between two atomic clouds \cite{SyncAtomicEnsemb,SyncExperiment}, in two spins subradiance \cite{leHur,GLG1,GLG2,Bellomo},
among optically pumped interacting dipoles \cite{SyncDipoles} and in trapped ions \cite{lee1,lee2,holland2018}.
A common key feature enabling quantum synchronization in these atomic systems is the presence 
of a collective dissipative coupling among atoms either because  this leads to a subradiant mode in relaxing systems 
or because superradiance allows overcoming other incoherent effects.

In this Letter, building on the proposed experimental scheme of Ref. \cite{giedke}, we design  a different setup, 
consisting of an atomic lattice in a dimer configuration,  where quantum simulation of 
SS can be realized. Atomic lattices represent a rich platform for many-body physics, entanglement and state engineering,
and for quantum simulation of condensed matter phenomena \cite{lattice}. Here we demonstrate the 
emergence of quan\-tum SS in atomic lattices even in the absence of collective dissipation, 
being instead the spatial modulation of the local decay rates the enabling factor. 
The phenomenon arises in dimer lattices and displays different mechanisms of SS, while it disappears in the limit of homogeneous chains,
similarly to synchronization blockade \cite{lorch,GLG2}. 
We also  compare local and global indicators of quantum 
SS in order to identify their relevance.

\textit{Dimer dissipative lattice.--}
One-dimensional optical lattices can be used to simulate an Ising-like dissipative spin chain, where spins are
 the two lower vibrational levels $|0\rangle$ and $|1\rangle$ of the atoms \cite{giedke}. 
 The system can be described as a two-band Bose-Hubbard model in the Mott-insulator regime \cite{Jaksch1}. Lattice anharmonicity, strong on-site repulsion,
 and perturbative  contributions due to weak tunneling among lattice sites,
lead to an $XXZ$ spin-$\frac{1}{2}$ chain Hamiltonian with highly adjustable parameters. 
Ising lattices can be simulated in a variety of platforms \cite{ising}, but the importance of the proposal \cite{giedke}  is the tunability of 
local Lindblad dissipation \cite{GKSL}, introduced by optically addressing the internal ($\Lambda$) structure of the atomic states
in the Lamb-Dicke regime.
The decay from the first excited motional state $|1\rangle$ towards $|0\rangle$ takes the usual form  
$\sum_{i}\gamma_i (2\hat{\sigma}_i^-\hat{\rho}\hat{\sigma}_i^+ -\{\hat{\sigma}_i^+\hat{\sigma}_i^-,\hat{\rho}\})$, 
where $i$ is the site index (see also Refs. \cite{marzoli,cirac}).   
Heating  can be experimentally made several orders of magnitude lower than cooling and then neglected\cite{giedke}.
\begin{figure}[t]
 \centering
 \includegraphics[width=0.9\columnwidth]{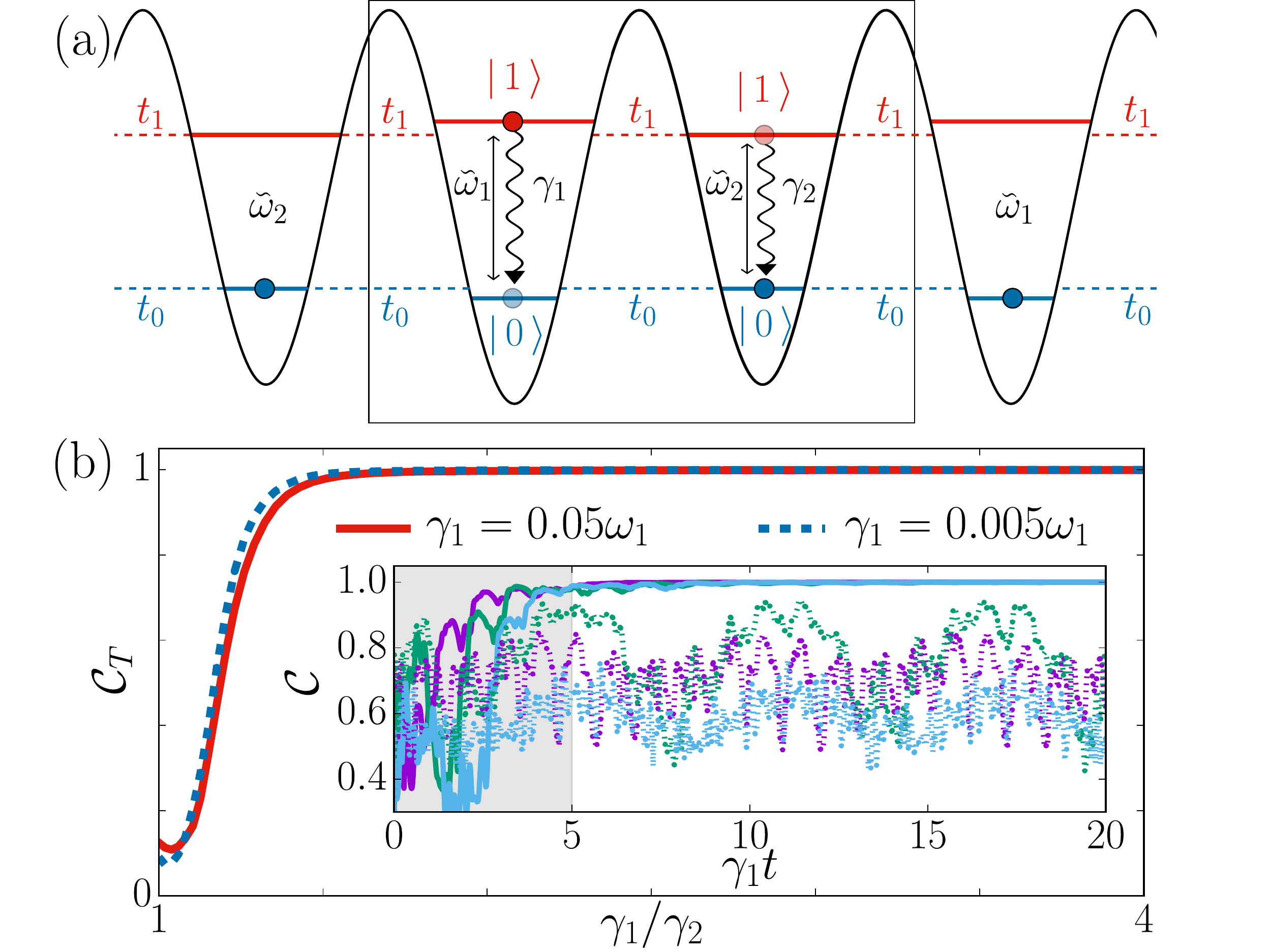}
 \caption{\label{fig1} (a)  
 Atomic lattice as described in main text. 
 (b) Emergent synchronization. Main panel: Global SS $\mathcal{C}_T(t=10/\gamma_1)$  as a function of $\gamma_1/\gamma_2$ for  two different 
 rates $\gamma_1$.  Inset:  Evolution of  $\mathcal{C}_{\langle \hat{\sigma}^x_j\rangle,\langle \hat{\sigma}^x_{j+1}\rangle}$ 
 for  $j=1$ (purple line), $j=2$ (in green), and $j=3$  (in blue),  with $\omega_1 \Delta t=3$. Here $\gamma_1/\omega_1=0.05$,
  solid lines correspond to $\gamma_1=4\gamma_2$ (staggered losses), dotted lines to $\gamma_1=\gamma_2$
  (homogeneous losses). 
  Shadowed region displays the transient after which the staggered case displays synchronization.
 System of  $N=4$ spins, $\delta=0.75\omega_1$, $\lambda=0.4\omega_1$, 
 initial state $|\Psi(t=0)\rangle=|+\rangle_1\otimes |+\rangle_2\otimes |+\rangle_3\otimes |+\rangle_4$, 
 where $|+\rangle_j=(|0\rangle+|\{1\}_j\rangle)/\sqrt{2}$ and $|\{1\}_j\rangle$ means spin $j$ excited, 
 the rest in vacuum. }
\end{figure}

Using standard techniques to produce double wells \cite{2well}, the dissipative model of  \cite{giedke} can be modified such that the lattice results in dimer arrangement 
where, 
in each well, the motional states have different (staggered) energy separation $\tilde{\omega}_{1,2}$ (Fig. \ref{fig1}a). Provided that the modulation of the optical wells can be treated as a perturbation, 
a dimer effective spin model can be obtained (further details in \cite{SupplementalMaterial})
with Hamiltonian ($\hbar=1$): 
\begin{equation}\label{Ham}
\hat{H}=\sum_{j=1}^{N}\frac{\omega_j}{2}\hat{\sigma}^z_j+ 
\sum_{j=1}^{N-1} \lambda(\hat{\sigma}^+_j\hat{\sigma}^-_{j+1}+h.c.), 
\end{equation}
where $\omega_j=\omega_{1(2)}$ if $j$ is odd (even), $\omega_j=\tilde{\omega}_j-\omega_0$, $\omega_0$ is the central large frequency  and
$\lambda$ is the spin-spin coupling. These parameters satisfy $\omega_0\gg\omega_{1,2},\delta,\lambda$, where $\delta=\omega_1-\omega_2$ is the detuning between the two sublattices.

The use of a bichromatic lattice will also affect the engineered dissipation, 
as the corresponding decay rates also depend  on the trap frequency through 
detuning with the cooling laser \cite{giedke}. 
In the weak dissipation regime, the reduced density matrix  $\hat{\rho}$ of the chain  obeys a standard 
 master equation \cite{GKSL}   
$\partial_t \hat{\rho}(t)=\mathcal{L}\hat{\rho}(t)$, with Liouvillian $\mathcal{L}\,\cdot=i[\cdot,\hat{H}]+\sum_j\gamma_j(2\hat{\sigma}^-_j\, \cdot\, \hat{\sigma}^+_j-\{\cdot,\hat{\sigma}^+_j\hat{\sigma}^-_j\})$. 
Because of the presence of a bichromatic lattice, the decay rates  $\gamma_j $ also assume staggered values 
and can be chosen such that $\gamma_1/\omega_1=\gamma_2/\omega_2$  \cite{SupplementalMaterial}.

A key observation for the  analysis of the dynamics is that the
whole eigenvalue spectrum of ${\cal L}$ can be analytically determined observing that it coincides with the one of   $\mathcal{K}$, which is defined through
 $\partial_t \hat{\rho}(t)=-i(\hat{K}\hat{\rho}(t)-\hat{\rho}(t)\hat{K}^\dagger)\equiv {\mathcal{K}} \hat{\rho}(t)$ and is 
obtained by replacing $\hat{H}$ with the non-Hermitian Hamiltonian 
$\hat{K}=\hat{H}-i\sum_j^N \gamma_j \hat{\sigma}^+_j\hat{\sigma}_j^-$ and by neglecting the jump operators  $\hat{\sigma}^-_j\,  \hat{\rho}\, \hat{\sigma}^+_j$.
In fact, in the eigenbasis of $\mathcal{K}$, $\mathcal{L}$ has an upper triangular form in which the diagonal elements are the eigenvalues of $\mathcal{K}$ (and then of $\mathcal{L}$  itself)  and the non-local jump operators only 
contribute to off-diagonal elements (see  Ref. \cite{Mauricio} for a detailed discussion). 

Diagonalization of $\hat{K}$ via Jordan-Wigner transformation and Fourier transform leads to the two-band elementary  complex eigenvalues \cite{SupplementalMaterial}  
\begin{equation}\label{Kspec}
\Omega^\pm_k=\frac{\Omega_1+\Omega_2}{2}\pm\frac{1}{2}\sqrt{(\Omega_1-\Omega_2)^2 +16\lambda^2 \,\text{cos}^2 \frac{k}{2} },
\end{equation}
and conjugates $\Omega_k^{\pm *}$ for $\hat{K}^\dagger$, where we assumed open boundary conditions, with $k=2\pi l/(N+1)$, $l=1,2,...,N/2$ and 
\begin{equation}
\Omega_{1(2)}=\omega_{1(2)}-i\gamma_{1(2)}. \label{link}
\end{equation}
The eigenvalues of $\mathcal{L}$ are obtained by combining $\Omega_k^{\pm}$ and $\Omega_k^{\pm *}$ as  prescribed in \cite{Mauricio}, such that their imaginary part (decay rates) always add together. Moreover, $-i\Omega_k^\pm$ and $i\Omega_k^{\pm*}$  are already eigenvalues of $\mathcal{L}$, corresponding to one-excitation eigenmodes.  Thus, the smallest decay rates of the system belong to this sector.  As a consequence, the relaxation dynamics before the final decay into the vacuum state 
is conveniently described  in the one-excitation sector, considering
the slowest modes that can be identified comparing their decay rates $\Gamma_l$ (absolute value of the imaginary parts of $\Omega_k^\pm$)
and frequencies $\nu_l$ (real parts of $\Omega_k^\pm$) \cite{ordering}.

\textit{Synchronization by staggered losses.--} 
We analyze the the full system dynamics  and quantify   the emergence of SS among 
atomic observables
with no classical counterpart, as the spin coherences  $\langle\hat{\sigma}^x_j\rangle$ \cite{ZambriniRev}. 
Indeed at any time during relaxation, coherences are present before reaching the equilibrium vacuum state. 
Their dynamical synchronization can be assessed by a Pearson correlation parameter $\mathcal{C}$ 
\cite{ZambriniRev}, a common measure of synchronization between  temporal trajectories, $x_{1(2)}(t)$,
defined as $\mathcal{C}_{x_1,x_2}(t)=\overline{\delta x_1 \delta x_2}/\sqrt{\overline{\delta x_1^2}\,\,
\overline{\delta x_2^2}}$, averaging on a time window $\Delta t$ of few oscillations 
$\overline{ x_j}=\frac{1}{\Delta t}\int_t^{t+\Delta t} ds\, x_j(s)$, 
and $\delta x_j=x_j -\overline{x_j}$.  
Delayed synchronization is accounted considering the correlation at
different times, $x_1(t)$ and $x_2(t+\tau)$, and maximizing over $\tau$, in general numerically, as for results presented in Fig. \ref{fig1}b. 
In Fig. \ref{fig1}b (inset) we show 
$\mathcal{C}_{\langle \hat{\sigma}^x_j\rangle,\langle\hat{\sigma}^x_{j'}\rangle}(t) $
among nearest-neighbor spin pairs, ranging between $0$ (no SS) to $1$ (perfect SS):
synchronization is found among all atoms  in the presence of local staggered dissipation
(solid lines), while it does  not emerge for $\gamma_1=\gamma_2$ (dotted lines). 
We also consider the global SS indicator 
 $\mathcal{C}_T(t)=\prod_{i<j}  \mathcal{C}_{\langle \hat{\sigma}^x_i\rangle,\langle 
\hat{\sigma}^x_j\rangle} (t)$   (main panel  of Fig. \ref{fig1}b) reaching its maximum value $1$ only 
if all atom coherences are synchronized. We see that SS 
for a given detuning ($\delta=0.75\omega_1$ in Fig. \ref{fig1}b) is enabled by the presence of staggered 
dissipation rates, emerging for a wide range of $\gamma_1/\gamma_2$ values, while it disappears if losses become uniform
($\gamma_1/\gamma_2\approx 1$).

The emergence of SS is due the presence of multiple dissipative time scales,  
as occurs in other models \cite{manzanoSciRep,GLG1,GLG2,Bellomo}.
Normal modes can conjure to dissipate at widely different rates, $\Gamma_l$, so that the predominant contribution to the long-time dynamics is represented by the slowest decaying mode.
Quantum SS then emerges as an ordered, spatially delocalized, monochromatic oscillation in the pre-asymptotic regime (transient synchronization). 
This is the case when considering our lattice with staggered dissipation, as revealed by inspection of the  Liouvillian spectrum. On the other hand, 
if local dissipation is  spatially homogeneous, the imaginary parts of the eigenvalues (\ref{Kspec}) coincide, 
there is no separation between decay rates, and in fact the system does not 
synchronize in spite of the presence of coherent coupling between spins (Fig. \ref{fig1}b).

\begin{figure}[h!]
 \centering
 \includegraphics[width=1\columnwidth]{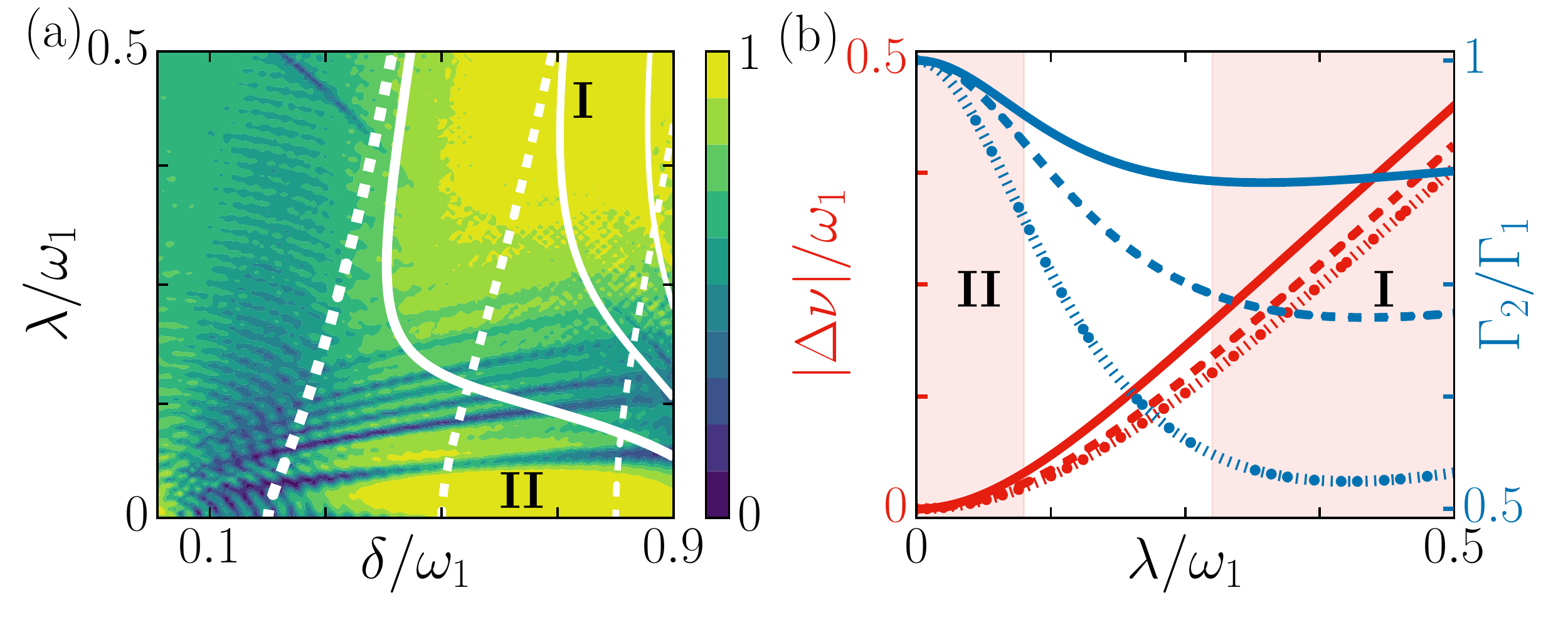}
 \caption{\label{syncmap} (Color online) (a) Map of SS among all spin pairs $\mathcal{C}_T(t)$
at $\gamma_1 t=10$ and $\omega_1 \Delta t=80$, varying detuning and coupling strength, 
 with $|\Psi(t=0)\rangle=|0\rangle/\sqrt{2}+(|\{1\}_2\rangle+|\{1\}_3\rangle)/2$. 
 Strong SS is found at the yellow (light color) regions I and II ($\mathcal{C}_T\ge0.9$).  White lines are contours 
 of the ratio of the two smallest decay rates, 
 $\Gamma_2/\Gamma_1$ (solid lines), and of the smallest decay rates of each band $\Gamma_2/\Gamma_3$  (dashed lines). 
 Thicker lines for increasing ratio values, 
 respectively $(0.9,0.75,0.6)$  and $(0.8,0.5,0.2)$. (b) In red, difference between the frequencies associated to the smallest decay 
 rates: $|\Delta \nu|=|\nu_1-\nu_2|$, 
 varying $\lambda$. In blue, ratio between the smallest decay rates. Solid lines $\delta=0.5\omega_1$, dashed lines 
 $\delta=0.75\omega_1$, dotted lines $\delta=0.9\omega_1$. For both figures $\gamma_j=0.05\omega_j$.}
\end{figure}

\textit{Inter-band and intra-band synchronization.--} 
The ability of the system to synchronize relies on the interplay between different parameters whose assessment can be conveniently
limited to the one-excitation sector of $\mathcal{L}$. Synchronization of the whole chain is calculated at a time long enough to wash out the presence of the slowest 
modes,  in Fig. \ref{syncmap}a, as a function of the  spin-spin coupling $\lambda$ and the detuning $\delta$, for a short chain of four spins. 
This SS map shows a non-trivial scenario with  
two different and well separated regions that support  SS, both occurring for strong detuning (yellow regions): region I characterized by strong coupling, and region II, 
by small coupling and a larger detuning window.

These SS regimes can be understood analyzing the  
spectral content of the coherences' dynamics   in the one-excitation 
sector $\langle 
\hat{\sigma}^x_j(t)\rangle=2Re\big[\sum_{l=1}^4 u_l(j)e^{-(i\nu_l+\Gamma_l)t}\big]$, 
with weights $u_l(j)$ depending on the site $j$, eigenmode $l$, and 
 initial condition \cite{SupplementalMaterial}. 
In region II the frequencies are nearly degenerate in each ($\pm$) band, while $N$ well separated 
frequencies are present in region I. 

The ``flat" and well separated two-band spectrum 
found in region II leads to what we term \textit{inter-band synchronization}. 
In fact, in the limit of vanishing $\lambda$,
the two frequency degenerate bands (Fig. \ref{syncmap}b) are separated by  $\delta+i (\gamma_1-\gamma_2)$. 
For weak coupling $\lambda$, two manifolds 
emerge with very similar frequencies and damping rates. Each of the sublattices of the atomic dimer is strongly coupled to 
one of the manifolds and weakly coupled to the other one, leading to an effective 
two-body behavior reminiscent of the mean-field scenario 
described in  \cite{SyncAtomicEnsemb,SyncDipoles}. Inter-band SS is present as long as the 
difference in local losses $\gamma_{1,2}$ allows one to establish two well separated time scales (region highlighted by white dashed lines in Fig. \ref{syncmap}a).
Being the staggered damping rates related to the sublattices detuning
(here we consider $\gamma_1/\omega_1=\gamma_2/\omega_2$),
SS only emerges for detuning $\delta$ larger than a threshold value,
at difference from the typical scenario of SS favored by small detuning \cite{ClassSync}
and similarly to synchronization blockade  \cite{lorch,GLG2}. This region shrinks when decreasing dissipation strength $\gamma_j$
as shown in \cite{SupplementalMaterial}.

Increasing the  coupling $\lambda$, SS deteriorates (Fig. \ref{syncmap}a, $0.1\lesssim \lambda/\omega_1\lesssim 0.25$)
as the two-body behavior disappears and several non-degenerate modes compete.
Synchronization is restored for coupling strengths such that there is a significant difference between the two slowest dissipation rates, now in the lower band,
so that a leading mode governs the long-time dynamics. This is \textit{intra-band synchronization} occurring in region I and requiring significant deviations 
between the slowest dissipation
rates (as highlighted by  white solid lines in Fig. \ref{syncmap}a). This picture  is confirmed when looking at the two slowest modes 
in Fig. \ref{syncmap}b, with frequencies and decay 
rates of the lower band drifting apart as the coupling increases.
 
When considering longer chains, the two physical mechanisms I and II for SS imply different levels of 
robustness.  In fact, inter-band synchronization II persists for long chains, as it mainly 
relies on the presence of the gap $\delta+i (\gamma_1-\gamma_2)$. This is not the case of region I, where 
the relevant spectral gap is obtained taking the difference between the two  
values of $\Omega^-_k$ with the smallest imaginary parts, which goes to zero as $N$ 
increases, Eq. (\ref{Kspec}). Furthermore, the simultaneous participation of all the eigenmodes of 
the lower band, makes the synchronized phase in region II almost spatially 
homogeneous, while the predominance of a single mode  in region I determines a 
nontrivial spatial structure, which also contributes to the loss of global synchronization  as size increases \cite{SupplementalMaterial}.
 
\begin{figure}[h!]
 \centering
 \includegraphics[width=1\columnwidth]{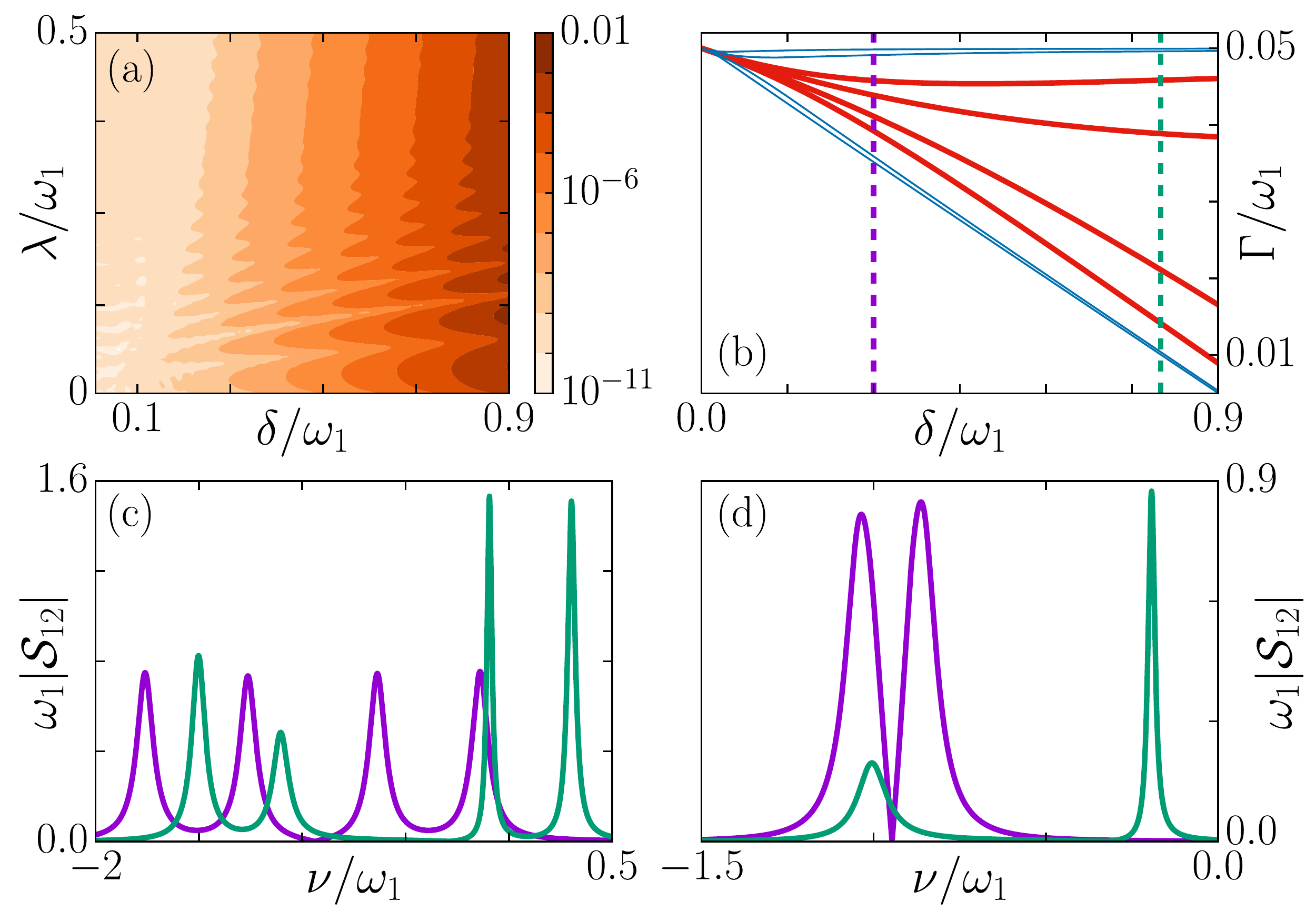}
 \caption{\label{correl}  (a) In color $|Z(\bar{t})|$, with $\gamma_1\bar{t}=10$, $|\Psi(t=0)\rangle=|0\rangle/\sqrt{2}+ \sum_{j=1}^4|\{1\}_j\rangle/\sqrt{8}$, and  averaged over few oscillations. We plot the absolute value as the magnitude of the correlations is the relevant quantity. (b) Decay rates $\Gamma_l$ as a function of the detuning, for $\lambda=0.5\omega_1$ (red) and $\lambda=0.05\omega_1$ (blue). (c,d) $\omega_1|\mathcal{S}_{12}(\nu/\omega_1)|$, with $\delta=0.1\omega_1$ (purple) and $\delta=0.8\omega_1$ (green). In (c) we fix $\lambda=0.5\omega_1$ (SS I), and in (d) $\lambda=0.05\omega_1$ (SS II). We plot the absolute values to ease peak comparison. In all plots  $\gamma_j/\omega_j=0.05$.}
\end{figure}

\textit{Synchronization measures.--} 
 Often, two-body quantum correlation indicators are taken as  \textit{bona fide} 
 synchronization measures \cite{ZambriniRev}, as they are able to reveal 
the presence of  phase locking. Here, we show that the presence 
of such correlations is necessary but not sufficient to predict the emergence of 
SS. We study the one-time correlation 
$Z( t )\equiv\langle\hat{\sigma}^x_1( t )\hat{\sigma}^x_{2}({t}
)\rangle$ often considered in the context of superradiance \cite{gross}, where $ t $ is set after the onset
of SS. As shown in Fig. \ref{correl}a, $Z$ 
increases with detuning (analogous results are found 
for other pairs) but it displays a weak dependence on $\lambda$,  being then unable to capture 
the transition from region I to region II.
An explanation can be given considering the one-excitation sector where  $Z( t )=2Re[\langle 
\hat{\sigma}^+_1( t )\hat{\sigma}^-_2( t ) \rangle 
]$ hence  $Z( t )= 2Re [\sum_{l,m=1}^4 
w_{l,m}(1,2)e^{[-i(\nu_l-\nu_m)-\Gamma_m-\Gamma_l] t }]$, with weights 
$w_{l,m}(1,2)$ depending on the spins, eigenmodes $l$ and $m$, and 
initial condition \cite{SupplementalMaterial}. The evolution of $Z$  is governed by exponentials 
containing  {\it combinations} of eigenvalues instead of single ones, depending then on slow and less slow rates. 
Therefore, differently from $\mathcal{C}$, it does not allow to distinguish the slowest relaxation modes.         

Different is the case for two-time correlation functions of the type $\langle \hat{\sigma}^-_{l}(t+\tau)\hat{\sigma}^+_m(t) \rangle$ 
in the stationary state, 
related to absorption spectra \cite{GambettaPRA2006} (for emission of radiating dipoles see Refs.\cite{SyncAtomicEnsemb,SyncDipoles}).
These are found to capture the presence of SS in both regimes I and II described above. 
In Fig. \ref{correl}c,d we plot 
$\mathcal{S}_{12}(\nu)=Re\big[\frac{1}{2\pi}\int_0^\infty d\tau e^{-i\nu 
\tau}\langle \hat{\sigma}^-_{1}(\tau)\hat{\sigma}^+_2(0) \rangle\big]$ 
in the (vacuum) stationary state of the system. In the 
one excitation sector we obtain $\langle 
\hat{\sigma}^-_1(\tau)\hat{\sigma}^+_2(0)\rangle=\sum_{l=1}^4 
v_{l}(1,2)e^{-(i\nu_l+\Gamma_l)\tau}$, with  weights $v_l(1,2)$  depending on 
the overlap of the eigenmode with the considered spin sites \cite{SupplementalMaterial}. We observe that the dynamics of these correlation functions is the same as the 
one for the spin coherences (with the initial condition $\hat{\sigma}_j^+|0\rangle$), 
in agreement with the quantum 
regression theorem \cite{Carmichael}. 
Thus the spectra $\mathcal{S}_{lm}(\nu)$ for each pair $l,m$ display a set  of at most $N$
resonance peaks, localized at the eigenfrequencies of the one-excitation 
sector, with linewidths determined by the corresponding decay rates, and height
depending on the weights $v_l(1,2)$.
This spectra contain  the information needed for the analysis of SS. 

In Fig. \ref{correl}c we plot a synchronized (green line) and an unsynchronized (purple line) two-time correlation function for strong coupling, while in Fig. \ref{correl}d we do it for weak coupling. In the absence of SS, for small detuning and strong coupling  (Fig. \ref{correl}c), 
the spectrum  displays multiple peaks, no one  
significantly sharper than the others.  For both small detuning and 
coupling, we find two 
peaks with similar decay rates (Fig. \ref{correl}d), while in the no-SS region 
between region I and II of Fig. \ref{syncmap}a two of the four peaks (the ones of 
the same band) display similar width \cite{SupplementalMaterial}.
Looking instead at SS parameter regions,  the 
spectra are characterized by the presence of a peak with width significantly 
smaller than the rest. 
Intra-band  synchronization (I) in Fig. \ref{correl}c displays
a  sharper third line among four, while in Fig. \ref{correl}d,  inter-band SS (II) clearly shows  
the effective two-body behavior of the system discussed above. This is also appreciated in Fig. \ref{correl}b where, for small coupling  (blue lines), two pairs of nearly degenerate decay rates emerge as detuning is increased. In contrast, for strong coupling (red lines), the system displays four well-differentiated decay rates. For both strong and weak coupling as detuning grows one of the peaks becomes significantly thinner, transiting from the purple spectra in Fig. \ref{correl}c,d to the green ones \cite{SupplementalMaterial}.  These   correlations, in the case of  {\it stationary} synchronization reported in Ref. \cite{SyncAtomicEnsemb}, are characterized by the presence of a single peak in the spectrum, a situation never occurring in our system for $\gamma_1/\gamma_2=\omega_1/\omega_2$.

 {\it Conclusions.--}   
We have shown  quantum spontaneous synchronization of an XX dissipative model in dimeric
spin chain that can be simulated in  atomic  lattices \cite{giedke}, and other set-ups.
Differently from other atomic systems, SS is enabled by spatial modulation of local losses, without any collective dissipation,
  and disappears in the homogeneous limit.  We have identified two SS regimes,
  interpreted in terms of the Liouvillian spectrum of the dynamics, and shown that inter-band SS is robust in long chains while intra-band 
  SS tends to disappear. We have analyzed the use of spin-spin correlations to assess the emergence of SS, comparing equal-time and two-time correlations. 
  While showing the limitations of the former, two-time correlation functions are found to properly demarcate the different regimes of transient (as well as stationary \cite{SyncAtomicEnsemb}) synchronization, looking at the number, 
  position and  width of 
  the peaks.
 Beyond the experimental realization of SS in atomic lattices as proposed here, future interesting directions are the generalization of the proposed atomic 
  set-up to display other forms of synchronization and the connection of this phenomenon with timely concepts such as time-crystals and coalescence.

  {\it  Acknowledgements.--}  The authors acknowledge support from the Horizon 2020 EU collaborative project
QuProCS (Grant Agreement No. 641277),  MINECO/AEI/FEDER through
projects EPheQuCS FIS2016-78010-P, CSIC Research Platform PTI-001, the Mar\'ia de Maeztu Program for Units of Excellence in R$\&$D (MDM-2017-0711),  
and funding from CAIB PhD and postdoctoral programs.

\clearpage
\begin{center}
\textbf{\large Supplemental Material: 'Quantum  synchronization in dimer atomic lattices'}
\end{center}
\setcounter{equation}{0}
\setcounter{figure}{0}
\setcounter{table}{0}
\setcounter{section}{0}
\makeatletter
\renewcommand{\thefigure}{S\arabic{figure}}
\def\theequation{S\arabic{equation}}
\renewcommand{\thesection}{S\arabic{section}}
\renewcommand*{\citenumfont}[1]{S#1}
\renewcommand*{\bibnumfmt}[1]{[S#1]}

\section{Dissipative atomic lattice}

In this section the main details on the physical implementation of the dissipative spin chain are overviewed. In \ref{S1A} we present the Hamiltonian that models the atomic lattice. We follow in \ref{S1B} explaining how from this atomic lattice one can realize effective spin Hamiltonians. In \ref{S1C} we comment on the proposed dissipation scheme of Ref. \cite{S_SpinDiss}, while we end in \ref{S1D} discussing briefly typical numerical values for the parameters of the atomic system.

\subsection{Two band Bose-Hubbard model}\label{S1A}

We consider a system of bosonic atoms in the Mott-insulator regime (MI), trapped in the two lowest energy bands of a bichromatic optical lattice. The optical lattice is assumed to be strongly anharmonic, such that higher vibrational levels are not  populated. The system is described by the following two-band Bose-Hubbard Hamiltonian:
\begin{equation}
\hat{H}_{BH}=\hat{H}_0+\hat{H}_1+\hat{H}_t. 
\end{equation}
The different contributions to this Hamiltonian are the following. $\hat{H}_0$ describes the atom-atom repulsive interactions and the unperturbed optical potential ($\hbar=1$) \cite{S_SpinDiss}:
\begin{equation}\label{H_0}
\begin{split}
\hat{H}_0=\sum_{j=1}^N \bigg\{\frac{\omega_0}{2}(\hat{d}^\dagger_j \hat{d}_j-\hat{c}^\dagger_j \hat{c}_j)+U_{01}\hat{c}^\dagger_j \hat{c}_j \hat{d}^\dagger_j \hat{d}_j\\
+\frac{U_{00}}{2}\hat{c}^\dagger_j \hat{c}_j(\hat{c}^\dagger_j \hat{c}_j-1)+\frac{U_{11}}{2}\hat{d}^\dagger_j \hat{d}_j(\hat{d}^\dagger_j \hat{d}_j-1)
\bigg\},
\end{split}
\end{equation}
$\hat{H}_1$ contains the small modulations to the optical potential:
\begin{equation}\label{H_1}
\hat{H}_1=\sum_{j\in \text{odd}}\frac{\omega_1}{2}(\hat{d}^\dagger_j \hat{d}_j-\hat{c}^\dagger_j \hat{c}_j)+
\sum_{j\in \text{even}}\frac{\omega_2}{2}(\hat{d}^\dagger_j \hat{d}_j-\hat{c}^\dagger_j \hat{c}_j),
\end{equation}
and $\hat{H}_t$ the perturbative tunneling processes between neighboring sites:
\begin{equation}\label{H_t}
\hat{H}_t=
\sum_{j=1}^{N-1}(t_0 \hat{c}^\dagger_j \hat{c}_{j+1}+t_1 \hat{d}^\dagger_j \hat{d}_{j+1}+H.c.). 
\end{equation}
The bosonic operators $\hat{c}_j^\dagger$ and $\hat{c}_j$ ($\hat{d}_j^\dagger$ and $\hat{d}_j$) create and annihilate an atom in the lowest (second lowest) motional state of site $j$ of the optical lattice. As in \cite{S_SpinDiss} the only atom-atom interactions are given by the same site and same motional state repulsion energies $U_{00}$ and $U_{11}$, and the same site different motional state repulsion $U_{01}$. Tunneling between neighboring sites without exchange of the motional state is permitted, with rates $t_0$ and $t_1$ \cite{S_SpinDiss}. Finally the motional states are separated by a large energy $\omega_0$ with small dimeric modulations $\omega_1$ and $\omega_2$. We consider the system to be in a regime in which there is one atom per site. We recall the usual hierarchy of parameter values that ensures the validity of the model \cite{S_Jaksch1,S_Jaksch2,S_SpinDiss}:
\begin{equation}\label{hierarchy}
\omega_0 \gg U_{00}, U_{11}, U_{01} \gg t_0,t_1 .
\end{equation}
Notice  that as we consider small frequency modulations of the optical potential, we must also require that:
\begin{equation}\label{hierarchy2}
\omega_0 \gg \omega_1,\omega_2, \quad U_{00},U_{11},U_{01}\gg \delta,
\end{equation}
with $\delta=\omega_1-\omega_2$. The first condition is necessary to be able to treat the modulations of the potential as a perturbation to the monochromatic Hamiltonian (\ref{H_0}).
The second additional condition is instead necessary to obtain the desired effective spin Hamiltonian (see next subsection).

\subsection{Effective spin Hamiltonian}\label{S1B}

An important observation is that $\hat{H}_{BH}$, besides conserving the total number of atoms $n$, also conserves  the total number of atoms in each motional state $n_0$ and $n_1$. Hence, the eigenstates of $\hat{H}_0+\hat{H}_1$ are given by the possible ways to distribute $n$ atoms in the two motional states of the optical lattice. Here we are interested in the low energy sector of the case $n=N$, in which there is one atom per site. In fact for prescribed values of $n_0$ and $n_1$, the lowest energy eigenstates, i.e. the ones with an atom per site, form a manifold of states with intra-energy separation of the order of $\delta$. In turn, all possible configurations in which there is one site with two atoms, form also a manifold with intra-energy separation again of order $\delta$. Both manifolds are separated by an energy gap of order of the repulsive interactions and hence much larger than the intra-manifold energy scales (Eq. (\ref{hierarchy2})). When considering $\hat{H}_t$, only matrix elements between 
unperturbed states of different manifold are non-zero. Then, if one is interested in the low energy physics of the system, one can use perturbation theory to obtain an effective Hamiltonian for the lowest energy manifold, and further neglect all states with more than one atom per site \cite{S_Cohen}. In this Schrieffer-Wolff kind of approach \cite{S_SpinDiss,S_Spins2,S_Spins3,S_Spins4}, second order tunneling processes couple the lowest energy states by means of virtual transitions to states with two atoms per site, which are energetically unfavorable. Thus to second order, one obtains the following effective Hamiltonian governing the manifold of states with one atom per site:
\begin{equation}\label{H_eff}
\begin{split}
\hat{H}_{\text{eff}}=\sum_{j=1}^{N-1} \bigg\{C_1\, \hat{c}_j^\dagger \hat{c}_j \hat{c}_{j+1}^\dagger \hat{c}_{j+1}+C_2\, \hat{d}_j^\dagger \hat{d}_j \hat{d}_{j+1}^\dagger \hat{d}_{j+1}\\
+C_3\,  (\hat{c}_j^\dagger \hat{c}_{j+1} \hat{d}_{j+1}^\dagger \hat{d}_{j}+\hat{d}_j^\dagger \hat{d}_{j+1} \hat{c}_{j+1}^\dagger \hat{c}_{j})\\
+C_4\, \hat{c}_j^\dagger \hat{c}_j \hat{d}_{j+1}^\dagger \hat{d}_{j+1}+ C_5\, \hat{d}_j^\dagger \hat{d}_j \hat{c}_{j+1}^\dagger \hat{c}_{j+1} \bigg\}\\
+\sum_{j\in \text{odd}} \big(\frac{\omega_0+\omega_1}{2} \big)(\hat{d}^\dagger_j \hat{d}_j-\hat{c}^\dagger_j \hat{c}_j)\\
+\sum_{j\in \text{even}}\big(\frac{\omega_0+\omega_2}{2} \big)(\hat{d}^\dagger_j \hat{d}_j-\hat{c}^\dagger_j \hat{c}_j),
\end{split}
\end{equation}
with the coefficients taking the following values:
\begin{equation}\label{coeff}
\begin{split}
C_1=-\frac{t_0^2}{U_{00}-\delta}-\frac{t_0^2}{U_{00}+\delta}\approx -\frac{2t_0^2}{U_{00}},\\
C_2=-\frac{t_1^2}{U_{11}-\delta}-\frac{t_1^2}{U_{11}+\delta}\approx-\frac{2t_1^2}{U_{11}},\\
C_3=-\frac{t_0 t_1}{U_{01}-\delta}-\frac{t_0 t_1}{U_{01}-\delta}\approx-\frac{2 t_0 t_1}{U_{01}},\\
C_4=-\frac{t_0^2+t_1^2}{U_{01}-\delta},\quad C_5=-\frac{t_1^2+t_0^2}{U_{01}+\delta},\\
C_4\approx C_5\approx -\frac{t_0^2+t_1^2}{U_{01}}.
\end{split}
\end{equation}
Notice that in (\ref{coeff}) we make use of the condition (\ref{hierarchy2}) to further approximate the expression of the coefficients. We can now define the spin states $\hat{c}^\dagger_j|0_j\rangle=|\downarrow_j\rangle$ and $\hat{d}^\dagger_j|0_j\rangle=|\uparrow_j\rangle$, together with the proper spin operators:
\begin{equation}
\begin{split}
\hat{\sigma}_j^+=\hat{d}_j^\dagger \hat{c}_j, \quad  \hat{\sigma}_j^-=\hat{c}_j^\dagger \hat{d}_j,\\
\hat{\sigma}_j^z=\hat{d}^\dagger_j \hat{d}_j- \hat{c}^\dagger_j \hat{c}_j,\\
\mathbb{1}_{2\times2}=\hat{d}^\dagger_j \hat{d}_j+ \hat{c}^\dagger_j \hat{c}_j,
\end{split}
\end{equation}
thus obtaining the following effective spin Hamiltonian:
\begin{equation}\label{HamSpin1}
\begin{split}
\hat{H}_{\text{spin}}=
\sum_{j=1}^{N-1} \big\{\lambda(\hat{\sigma}_j^+ \hat{\sigma}_{j+1}^- +\hat{\sigma}_{j+1}^+ \hat{\sigma}_{j}^- )+\lambda_z \hat{\sigma}_j^z\hat{\sigma}_{j+1}^z \big\}\\
+\sum_{j\in\text{odd}}\bigg(\frac{\omega_0+\omega_1}{2}+h_z\bigg)\hat{\sigma}_j^z+ \sum_{j\in \text{even}}\bigg(\frac{\omega_0+\omega_2}{2}+h_z\bigg)\hat{\sigma}_j^z.
\end{split} 
\end{equation}
with the parameters defined as:
\begin{equation}\label{param_eff}
\begin{split}
\lambda\approx -\frac{2t_0 t_1}{U_{01}},\quad h_z\approx\frac{1}{2}\bigg(\frac{t_0^2}{U_{00}}-\frac{t_1^2}{U_{11}} \bigg),\\
\lambda_z\approx -\frac{1}{2}\bigg(\frac{t_0^2}{U_{00}}+\frac{t_1^2}{U_{11}} -\frac{t_0^2+t_1^2}{U_{01}}\bigg).
\end{split}
\end{equation}
Finally,  $\hat{H}$ of Eq. (1) in the main text corresponds to parameters of the optical lattice such that $\lambda_z=0$. Then $\hat{H}$ is $\hat{H}_{\text{spin}}$ in a frame rotating with $\frac{\omega_0}{2}+h_z$.

\subsection{Engineered dissipation}\label{S1C}

A detailed derivation of the dissipation scheme used in this work is found in Ref. \cite{S_SpinDiss} and here we review the main conditions to implement it. 
It is assumed that the atoms are in the Lamb-Dicke regime ($\eta_j\ll1$, where $\eta_j$ is the Lamb-Dicke parameter at site $j$) and have a `$\Lambda$' internal structure with two ground states. By means of weak off-resonant Raman transitions the excited state is adiabatically 
eliminated, leading to an effective two level system with tunable decay rates \cite{S_SpinDiss,S_cooling2,S_cooling3}. This effective two-level system is characterized by an effective Rabi frequency 
$\Omega_{\text{eff}}$, an effective detuning $\delta_r$, an effective decay rate $\Gamma$, and an effective dephasing rate $\gamma$, where the expression for these parameters is found in many references
 \cite{S_SpinDiss,S_cooling2,S_cooling3}. The parameters are then adjusted so that the two-level system resolves the motional degrees of freedom, i.e. $\Gamma+\gamma\ll\tilde{\omega}_j$ 
(with $\tilde{\omega}_j=\omega_0+\omega_j$) \cite{S_SpinDiss,S_cooling2,S_cooling3}. Finally, 
if $\eta_j|\Omega_{\text{eff}}|\ll\Gamma,\gamma,|\delta_r|,\tilde{\omega}_j$ \cite{S_SpinDiss}, the parameter regime is characterized by weak coupling of internal and motional degrees of freedom, and 
one can adiabatically eliminate the former obtaining an effective master equation for the motional degrees of freedom \cite{S_cooling1}. Under the appropriate resonance conditions, heating can be neglected, 
and one obtains that the density matrix evolves according to the Liouvillian $\mathcal{L}\,\cdot=i[\cdot,\hat{H}]+\sum_j\gamma_j(2\hat{\sigma}^-_j\, \cdot\, \hat{\sigma}^+_j-\{\cdot,\hat{\sigma}^+_j\hat{\sigma}^-_j\})$ with motional decay rate:
\begin{equation}
\gamma_j=\eta_j^2\Omega_{\text{eff}}^2\frac{\Gamma+\gamma}{(\Gamma+\gamma)^2+(\delta_r-\tilde{\omega}_j)^2}. 
\end{equation}
Notice that, in order to suppress heating, the effective detuning should be tuned close to the large mechanical energy, i.e. $\delta_r\sim\omega_0$. Then the dependence of the decay rate on the lattice site comes mainly from the resonance frequency of the Lorentzian, as differences between $\eta_1^2$ and $\eta_2^2$ are of the order of $\omega_j/\omega_0\ll1$.  Defining $\epsilon=\delta_r-\omega_0\sim|\omega_j|$ (which can be positive or negative), and by properly adjusting it, the decay rates can in general take staggered values. Furthermore, besides implementing staggered decay rates by means of the effective detuning (as described here), different approaches are also proposed  in \cite{S_SpinDiss}, as for example by tuning the phase difference between two cooling lasers.

\subsection{Brief survey of parameter values}\label{S1D}

We follow references \cite{S_Jaksch1,S_Jaksch2} to illustrate the values that the parameters of this system can take. For sodium atoms in a blue detuned optical trap of wavelength $\lambda_T=514$nm, the recoil energy is $E_R=2\pi\times33$kHz. Tuning the light intensity, the energy separation between the two lowest energy motional states of the lattice can be fixed to $\omega_0\sim1$MHz, which leads to $U_{lm}\sim40$kHz and $t_l\sim4$kHz ($l,m=0,1$). In these conditions the atomic chain is in the MI regime with one atom per site. Moreover, according to eq. \ref{param_eff}, $\lambda\sim0.8$kHz which sets the order of magnitude of the small modulations $\omega_j$, as we take $\omega_j\sim\lambda$ in all the work. Considering possible sources of dissipation, we notice that in the MI regime with one atom per site atom-atom collisions are strongly suppressed \cite{S_Jaksch1}. In addition, for these parameter values, the rate of dissipation due to the optical potential can be estimated to be of the order of $10^{-2}$Hz \cite{S_Jaksch2}, and we neglect it. This last approximation is consistent with the much larger values that we have fixed for the engineered decay rates, $\gamma_j$, which we estimate to be in the range $\sim 1-40$Hz.

\section{Liouvillian spectrum}

As it is shown in Ref. \cite{S_Mauricio}, the eigenvalues of the type of Liouvillian $\mathcal{L}$ considered  here, are prescribed linear combinations of those of the non-Hermitian 
Hamiltonian $\hat{K}=\hat{H}-i\sum_j^N \gamma_j \hat{\sigma}^+_j\hat{\sigma}_j^-$. In fact, as commented in the main text, the eigenvalues with the smallest decay rates coincide with eigenvalues of 
$\hat{K}$ and their 
complex conjugates. Thus in order to characterize the long-time relaxation dynamics we need to diagonalize $\hat{K}$. To do so we use the Jordan-Wigner 
transformation to work with fermions instead of spins:
\begin{equation}\label{JW}
\begin{split}
\hat{\sigma}^z_j=2\hat{f}^\dagger_j \hat{f}_j-1,\\
\hat{\sigma}^+_j=\hat{f}^\dagger_j e^{i\hat{\phi}_j}, \quad \hat{\sigma}^-_j=\hat{f}_j e^{-i\hat{\phi}_j},\\
\hat{\phi}_j=\pi\sum_{l<j} \hat{n}_l,
\end{split} 
\end{equation}
where $\hat{f}_j^\dagger (\hat{f}_j)$, $\hat{n}_j=\hat{f}_j^\dagger \hat{f}_j$,  are the creation (annihilation) and number fermionic operators of site $j$. 
Then defining $\Omega_{1(2)}=\omega_{1(2)}-i\gamma_{1(2)}$, and using a notation that displays explicitly the dimeric character of the chain, we obtain the {\it fermionic} non-Hermitian Hamiltonian:
\begin{equation}\label{K_F}
\begin{split}
\hat{K}_F=\sum_{j=1}^{N/2}\Omega_1 \hat{a}^\dagger_j \hat{a}_j +\sum_{j=1}^{N/2}\Omega_2 \hat{b}^\dagger_j \hat{b}_j\\
+ \sum_{j=2}^{N/2} \lambda(\hat{a}^\dagger_j \hat{b}_{j-1}+H.c)+\sum_{j=1}^{N/2}\lambda(\hat{a}^\dagger_j \hat{b}_j+H.c.),
\end{split}
\end{equation}
where now $\hat{a}_j^\dagger(\hat{a}_j)$ and $\hat{b}_j^\dagger(\hat{b}_j)$ are fermionic creation (annihilation) operators of site $j$ and its basis, respectively. The diagonalization of $\hat{K}_F$ is accomplished in two steps: first we diagonalize the system assuming periodic boundary conditions, and later we combine the obtained eigenstates to find the ones of the open boundary case. In the following we write the main results of each step.

\subsection{Periodic boundary conditions}

In this case the summation in the third term of Eq. (\ref{K_F}) starts from $j=1$, and the boundary conditions imply that $\hat{a}_{0}=\hat{a}_{N/2}$ and  $\hat{b}_{0}=\hat{b}_{N/2}$. 
We define $M=N/2$ and relabel the index $j$ to run from $j=0$ to $M-1$. Then exploiting translational invariance we define the Fourier modes (denoted by $k$ index):
\begin{equation}
\begin{split}
\hat{a}_j=\frac{1}{\sqrt{M}}\sum_k \hat{a}_k e^{-i\frac{k}{2}}e^{ikj},\\
\hat{b}_j=\frac{1}{\sqrt{M}}\sum_k \hat{b}_k e^{ikj},
\end{split} 
\end{equation}
with $k=2\pi l/M$ and $l=0,1,\dots,M-1$. Notice that we have anticipated the need of a complex phase $-k/2$ in the expression for the $\hat{a}_j$'s. These modes leave $\hat{K}_F$ in a block-diagonal form $\hat{K}_F=\bigoplus_k\hat{K}_F(k)$ in which each block is given by:
\begin{equation}\label{K_blocks}
\hat{K}_F(k)= (\hat{a}^\dagger_k \,\,\, \hat{b}^\dagger_k) 
 \begin{pmatrix}
    \Omega_1 & 2\lambda\cos \frac{k}{2}\\
    2\lambda\cos \frac{k}{2} & \Omega_2
 \end{pmatrix}
  \begin{pmatrix}
    \hat{a}_k\\
    \hat{b}_k
 \end{pmatrix}.
\end{equation}
A sufficient but not necessary condition for this non-Hermitian matrix to be diagonalizable is that it is not degenerate. Note that this is always fulfilled in the parameter region $\omega_1\neq\omega_2$ and $\omega_{1(2)}>\gamma_{1(2)}$. The diagonalization is accomplished by means of an orthogonal transformation defined as:
\begin{equation}
\begin{split}
\hat{\alpha}_k(\hat{\alpha}'_k)=\hat{a}_k(\hat{a}^\dagger_k)\cos \theta_k -\hat{b}_k(\hat{b}^\dagger_k)\sin \theta_k ,\\
\hat{\beta}_k(\hat{\beta}'_k)=\hat{a}_k(\hat{a}^\dagger_k)\sin \theta_k +\hat{b}_k(\hat{b}^\dagger_k)\cos \theta_k , 
\end{split} 
\end{equation}
with 
\begin{equation}
\tan 2\theta_k= -\frac{4\lambda \cos \frac{k}{2}}{\Omega_1-\Omega_2}.
\end{equation}
Importantly as $\theta_k$ is complex, this orthogonal transformation is not unitary and hence $\hat{\alpha}'_k(\hat{\beta}'_k)\neq \hat{\alpha}^\dagger_k(\hat{\beta}^\dagger_k)$. Only in the case $\gamma_1=\gamma_2=0$, $\theta_k$ becomes real and we recover the standard operators. The eigenvalues of (\ref{K_blocks}) are given by:
\begin{equation}\label{Kspec}
\Omega^\pm_{k}=\frac{\Omega_1+\Omega_2}{2}\pm\frac{1}{2}\sqrt{(\Omega_1-\Omega_2)^2 +16\lambda^2 \,\text{cos}^2 \frac{k}{2} },
\end{equation}
with the $k$'s as above prescribed, and the correspondence of band '$+(-)$' to operator $\hat{\alpha}_k(\hat{\beta}_k)$.  An important characteristic of this spectrum is that under the transformation $l\to M-l$ yields $\Omega^\pm_{k_l}=\Omega^\pm_{k_{M-l}}$ and $\theta_{k_l}=-\theta_{k_{M-l}}$. Indeed, part of the Fourier modes appear in pairs of degenerate eigenvalues, here corresponding  to the pairs with the {\it k}'s associated to $\{l,M-l\}$. Besides the degenerate eigenmodes, there is  the mode $k=0$, and when $M$ is even there is also $k=\pi$. Notice that, although  the spectrum is partially degenerate, the Fourier modes for different $k$ are linearly independent and hence the set of eigenvectors too, as it is required for a  matrix to be diagonalizable.  Finally we write down the expression of $\hat{\alpha}_k$ and $\hat{\beta}_k$ in the site basis:
\begin{equation}\label{vectors_pbc}
\begin{split}
\hat{\alpha}_k=\frac{1}{\sqrt{M}}\sum_{j=0}^{M-1}\big(\hat{a}_j\cos \theta_k e^{i\frac{k}{2}}-\hat{b}_j\sin \theta_k \big)e^{-ikj}, \\
\hat{\beta}_k=\frac{1}{\sqrt{M}}\sum_{j=0}^{M-1}\big(\hat{a}_j\sin \theta_k e^{i\frac{k}{2}}+\hat{b}_j\cos \theta_k \big)e^{-ikj}. 
\end{split} 
\end{equation}
\subsection{Open boundary conditions}

In this case, we first consider a larger system of $M'=2M+1$ cells with periodic boundary conditions and we take linear combinations of its degenerate eigenmodes, i.e.
$\hat{u}_{k_l}=x_1\hat{\alpha}_{k_l}+x_2\hat{\alpha}_{k_{M'-l}}$ and $\hat{v}_{k_l}=y_1\hat{\beta}_{k_l}+y_2\hat{\beta}_{k_{M'-l}}$, with $l=1,2, ..., M$. By requiring $\hat{u}_{k_l}(\hat{v}_{k_l})$ to be zero at sites $b_0$ and $a_{M+1}$,
we can  obtain the eigenmodes of the open boundary case with $M$ cells. In particular the first condition is satisfied for any $k$ if we take $x_1=x_2$ and $y_1=-y_2$, i.e. we replace as usual the exponentials by sines. 
Then we see that the sine modes have a vanishing amplitude on $a_{M+1}$ too, as it follows from the definition of the allowed $k$'s:
\begin{equation}\label{k_obc}
 k=\frac{2\pi l}{N+1}\,\,\implies\,\,\sin[k(M+\frac{1}{2})]=0, 
\end{equation}
with $N=2M$. Hence the normalized eigenmodes read as:
\begin{equation}\label{vectors_obc}
\begin{split}
\hat{u}_k=\sqrt{\frac{4}{N+1}}\sum_{j=1}^{M}\big(\hat{a}_j\cos\theta_k \sin\,[k(j-\frac{1}{2})]\\
- \hat{b}_j\sin\,\theta_k \sin\,[k j]\big),\\
\hat{v}_k=\sqrt{\frac{4}{N+1}}\sum_{j=1}^{M}\big(\hat{a}_j\sin\theta_k \sin\,[k(j-\frac{1}{2})]\\
+ \hat{b}_j\cos\,\theta_k \sin\,[k j]\big),
\end{split} 
\end{equation}
where the eigenvalues of $\hat{u}_k(\hat{v}_k)$'s belong to the '$+(-)$' band. We can obtain $\hat{u}'_k(\hat{v}'_k)$  
by replacing the operators $\hat{a}_j(\hat{b}_j)$ by  $\hat{a}^\dagger_j(\hat{b}^\dagger_j)$. Again, $\hat{u}'_k(\hat{v}'_k)\neq\hat{u}^\dagger_k(\hat{v}^\dagger_k)$, 
except for the case $\gamma_{1(2)}=0$, for the same reasons as before. Notice that this  set of eigenvectors forms a complete orthogonal basis, both for $\theta_k$ real and complex.

\section{Dynamics in the one-excitation sector}

In the one excitation sector, the phase $\hat{\phi}_j$ of the Jordan-Wigner transformation (\ref{JW}) is zero. Then fermionic and spin operators are equivalent, 
and the master equation in the fermionic picture takes the same form as the spin one. $\hat{f}_j(\hat{f}^\dagger_j)$ denote the fermionic annihilation (creation) operators, 
which in the one excitation sector are equivalent to the spin coherences.  Moreover, it is useful to use the following notation. In the site basis we define $\hat{f}_j^\dagger|0\rangle=|F_j\rangle$, and $\langle 0|\hat{f}_j=\langle F_j|$,
while we use eqs. (\ref{vectors_obc}) to define $\hat{u}'_k(\hat{v}'_k)|0\rangle=|{K}_l\rangle$, and $\langle 0|\hat{u}_k(\hat{v}_k)=\langle {K}^*_l|$, with $l$ running from 1 to N and the first half belonging to the energy band '$-$', while the other to the '$+$' band (as in the main text). $|{K}_l\rangle$, $\langle {K}^*_l|$ correspond to the right and left eigenvectors of $\hat{K}_F$ respectively. Notice that, as $\hat{K}_F$ is represented by a non-Hermitian symmetric matrix, the left eigenvectors are just the transpose of the right ones, as the '*' indicates in the bra-ket notation. Moreover,  $\langle {K}^*_l|{K}_{l'}\rangle=\delta_{l,l'}$.

\subsection{Exact time evolution}

We now rewrite the master equation in the fermionic basis as $\partial_t \hat{\rho}=-i(\hat{K}_F\hat{\rho}-\hat{\rho}\hat{K}^\dagger_F)+2\sum_j\gamma_j\hat{f}_j\hat{\rho}\hat{f}_j^\dagger$.  Notice that in the one-excitation sector only density matrix terms of the type $|F_j\rangle\langle F_{j'}|$, $|F_j\rangle\langle 0|$ and $|0\rangle\langle F_{j}|$ contribute to the expectation values we are interested in. Moreover, in the one excitation sector, the jump part of the master equation does not contribute to the time evolution of these quantities. Thus we only need to consider $\partial_t \hat{\rho}=\hat{\mathcal{K}}\hat{\rho}$, with $\hat{\mathcal{K}}\hat{\rho}=-i(\hat{K}_F\hat{\rho}-\hat{\rho}\hat{K}^\dagger_F)$. As  $\hat{K}^\dagger_F$ is $\hat{K}_F$ with $\Omega^*_{1(2)}$, its eigenvalues and eigenstates are obtained from Eq. (\ref{Kspec}) and (\ref{vectors_obc}) making the same replacement. Thus the right and left eigenvectors of $\hat{K}^\dagger_F$ are $|K^*_l\rangle$ and $\langle K_l|$ 
respectively. Taking all these into account, we can write:
\begin{equation}
\begin{split}
\hat{\mathcal{K}}|{K}_l\rangle\langle{K}_m|&=[-i(\nu_l-\nu_m)-\Gamma_l-\Gamma_m] |{K}_l\rangle\langle{K}_m|,\\
\hat{\mathcal{K}}|{K}_l\rangle\langle 0|&=-(i\nu_l+\Gamma_l) |{K}_l\rangle\langle0|,\\
\hat{\mathcal{K}}|0\rangle\langle {K}_m|&=(i\nu_m-\Gamma_m) |0\rangle\langle {K}_m|.\\
\end{split}
\end{equation}
Defining the following projectors:
\begin{equation}\label{proj_nH}
\begin{split}
{\mathcal{P}}_{l,m}\hat{\rho}(t)&= (\langle{K}^*_l|\hat{\rho}(t) |{K}^*_m\rangle)|{K}_l\rangle\langle{K}_m|,\\ 
{\mathcal{P}}_{l,0}\hat{\rho}(t)&= (\langle{K}^*_l|\hat{\rho}(t) |0\rangle)|{K}_l\rangle\langle0|,\\
{\mathcal{P}}_{0,m}\hat{\rho}(t)&= (\langle0|\hat{\rho}(t) |{K}^*_m\rangle)|0\rangle\langle{K}_m|,
\end{split}
\end{equation}
we can write the time evolution of these density matrix projections as:
\begin{equation}\label{evol_nH}
\begin{split}
{\mathcal{P}}_{l,m}\hat{\rho}(t)&={\mathcal{P}}_{l,m}\hat{\rho}(0)e^{[-i(\nu_l-\nu_m)-\Gamma_l-\Gamma_m]t} ,\\ 
{\mathcal{P}}_{l,0}\hat{\rho}(t)&={\mathcal{P}}_{l,0}\hat{\rho}(0)e^{-(i\nu_l+\Gamma_l)t} ,\\
{\mathcal{P}}_{0,m}\hat{\rho}(t)&={\mathcal{P}}_{0,m}\hat{\rho}(0)e^{(i\nu_m-\Gamma_m)t}.
\end{split}
\end{equation}
Finally notice that the explicit form of the Liouvillian eigenmodes can be found generalizing the two-spin results of Ref. \cite{S_Bellomo} to $k$-dependent couplings.

\subsection{Main results}

\begin{figure}[H]
 \centering
 \includegraphics[width=0.95\columnwidth]{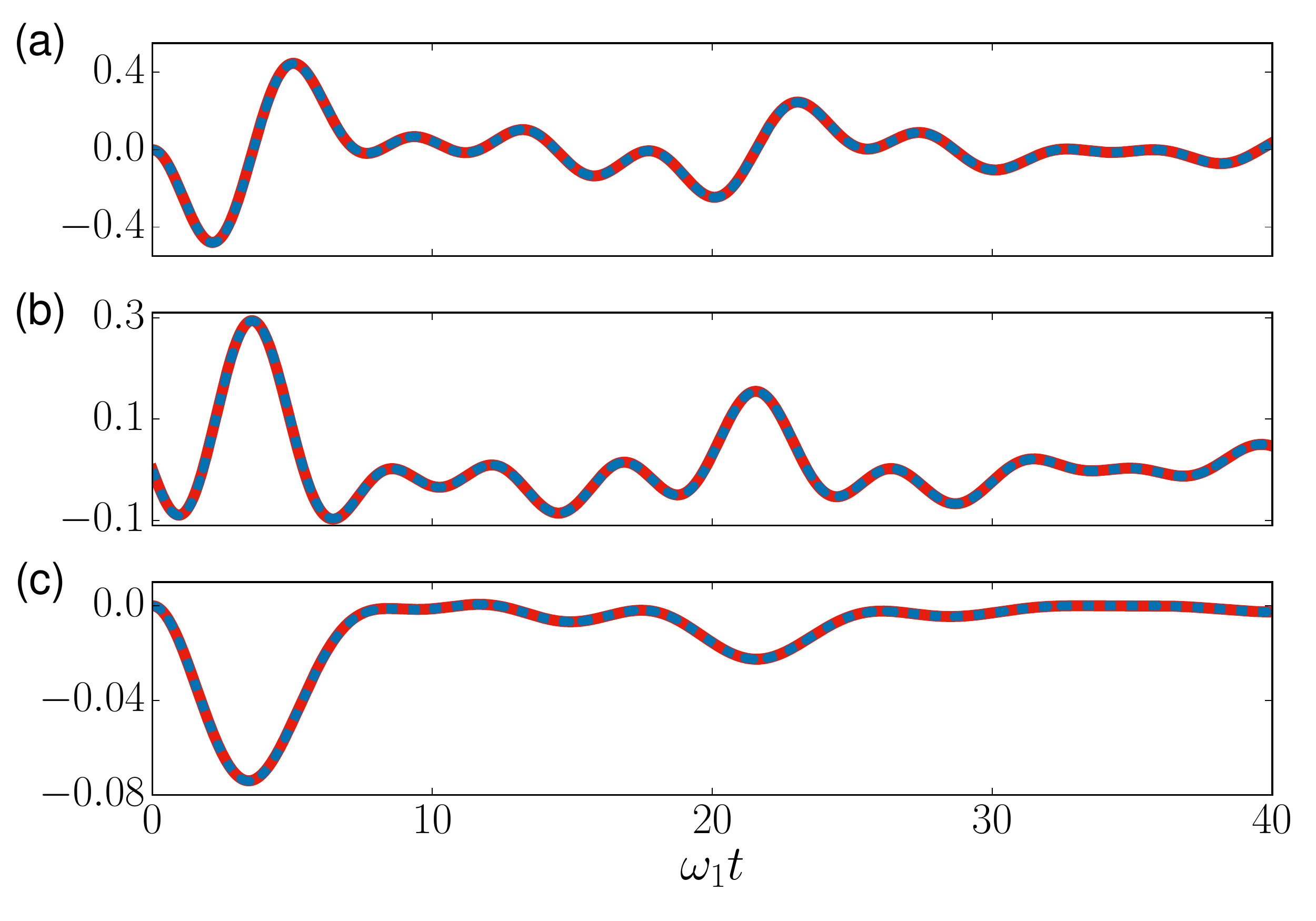}
 \caption{\label{figureS1} In all cases the parameters are fixed to $N=4$, $\delta=0.25\omega_1$, $\lambda=0.3\omega_1$ and $\gamma_j/\omega_j=0.05$. Red solid lines correspond to exact numerical results, blue dashed lines to the analytical expressions derived in this section. (a) $\langle \hat{\sigma}^x_1(t)\rangle$, considering the initial condition $|\Psi_0\rangle=(|0\rangle+|F_2\rangle)/\sqrt{2}$. (b) Imaginary part of $\langle \hat{\sigma}^-_1(t)\hat{\sigma}^+_2(0)\rangle$. Notice that the real part of this quantity is the same as (a). (c)  $\langle \hat{\sigma}^x_1(t)\hat{\sigma}^x_2(t)\rangle$, considering the initial condition $|\Psi_0\rangle=(|0\rangle+|F_2\rangle)/\sqrt{2}$.}
\end{figure}

We first write down the fermionic operators in the following way:
\begin{equation}
\hat{f}_j=|0\rangle\langle F_j|,\quad
\hat{f}^\dagger_j=|F_j\rangle\langle 0|,\quad f^\dagger_jf_{j'}=|F_j\rangle\langle F_{j'}|.
\end{equation}
Then using these definitions and Eqs. (\ref{proj_nH}) and (\ref{evol_nH}), we can obtain the expressions for the time evolution of the expected values presented in the main text. First we consider $\langle \hat{\sigma}^x_j(t)\rangle$, with an initial condition $\hat{\rho}(0)=|\Psi_0\rangle\langle \Psi_0|$. Hence:
\begin{equation}\label{sigma_x_t}
\begin{split}
\langle \hat{\sigma}^x_j(t)\rangle=2Re\big( \text{Tr}[\hat{f}_j\hat{\rho}(t)]\big)=\\
=2Re \big(\text{Tr}[\hat{f}_j\sum_l\mathcal{P}_{l,0}\hat{\rho}(t)]\big), 
\end{split}
\end{equation}
which yields:
\begin{equation}
\langle \hat{\sigma}^x_j(t)\rangle=2Re\big(\sum_l u_l(j) e^{-(i\nu_l+\Gamma_l)t}  \big), 
\end{equation}
with
\begin{equation}
u_l(j)=\langle {K}^*_l|\Psi_0\rangle\langle\Psi_0|0\rangle\langle F_j| {K}_l\rangle. 
\end{equation}
Next we consider the correlation $\langle \hat{\sigma}^x_j(t)\hat{\sigma}^x_{j'}(t)\rangle$, which in the one excitation picture is given by $2Re[\langle \hat{f}^\dagger_j(t) \hat{f}_{j'}(t)\rangle]$. Proceeding analogously, we find that:
\begin{equation}
\text{Tr}[\hat{f}^\dagger_j \hat{f}_{j'}\hat{\rho}(t)]=\text{Tr}[\hat{f}^\dagger_j \hat{f}_{j'}\sum_{l,m}\mathcal{P}_{l,m}\hat{\rho}(t)], 
\end{equation}
and hence:
\begin{equation}
\begin{split}
&\langle \hat{\sigma}^x_j(t)\hat{\sigma}^x_{j'}(t)\rangle=\\
&=2Re\big(\sum_{l,m}w_{l,m}(j,j') e^{[-i(\nu_l-\nu_m)-\Gamma_l-\Gamma_m]t} \big),
\end{split}
\end{equation}
with:
\begin{equation}
w_{l,m}(j,j')=\langle {K}^*_l|\Psi_0\rangle\langle \Psi_0 | {K}^*_m\rangle\langle {K}_m|F_j\rangle \langle F_{j'}|{K}_l\rangle. 
\end{equation}
Finally we consider the two time correlation function $\langle \hat{\sigma}^-_j(\tau)\hat{\sigma}^+_{j'}(0)\rangle$ where $0$ denotes an arbitrary time origin in the stationary state (the vacuum). With the help of quantum regression theorem \cite{S_Carmichael}, we know that this is equivalent to compute the time evolution of $\langle \hat{\sigma}^-_j(\tau)\rangle$ with the initial condition $\hat{\sigma}^+_{j'}|0\rangle\langle0|$. Thus proceeding analogously as for (\ref{sigma_x_t}), we obtain:
\begin{equation}
\langle \hat{\sigma}^-_j(\tau)\hat{\sigma}^+_{j'}(0)\rangle= \sum_l v_l(j,j')  e^{-(i\nu_l+\Gamma_l)\tau}, 
\end{equation}
with
\begin{equation}
v_l(j,j')=\langle {K}^*_l|F_{j'}\rangle\langle F_j| {K}_l\rangle. 
\end{equation}
We also write down the Fourier transform of this correlation $\mathcal{S}_{jj'}(\nu)$ studied in the main text:
\begin{equation}\label{spectrum_eq}
\begin{split}
\mathcal{S}_{jj'}(\nu)=Re\big[\frac{1}{2\pi}\int_0^\infty d\tau e^{-i\nu 
\tau}\langle \hat{\sigma}^-_{j}(\tau)\hat{\sigma}^+_{j'}(0) \rangle\big]\\
=\frac{1}{2\pi}\sum_l \frac{\Gamma_l Re[v_l(j,j')]+(\nu+\nu_l)Im[v_l(j,j')]}{\Gamma_l^2+(\nu+\nu_l)^2}.
\end{split}
\end{equation}
These equations are the results we use in the main text to compare and analyze different synchronization measures. In Fig. \ref{figureS1} we plot an example for each of these quantities, comparing  numerical trajectories with the  analytical results as a consistency check, finding that they agree. 

\section{Additional results about synchronization}

We present further results that complement the discussion on the emergence of synchronization of the main text. In particular we show an example of a synchronized trajectory for region I and II \ref{sect4a}, we give more details on the effect of varying spin's dissipation strength \ref{sect4b}, and on varying the size of the chain \ref{sect4c}, and  we show how the two-time correlation functions change with detuning and coupling \ref{sect4d}. 

\subsection{Synchronized trajectories}\label{sect4a}

\begin{figure}[H]
 \centering
 \includegraphics[width=0.95\columnwidth]{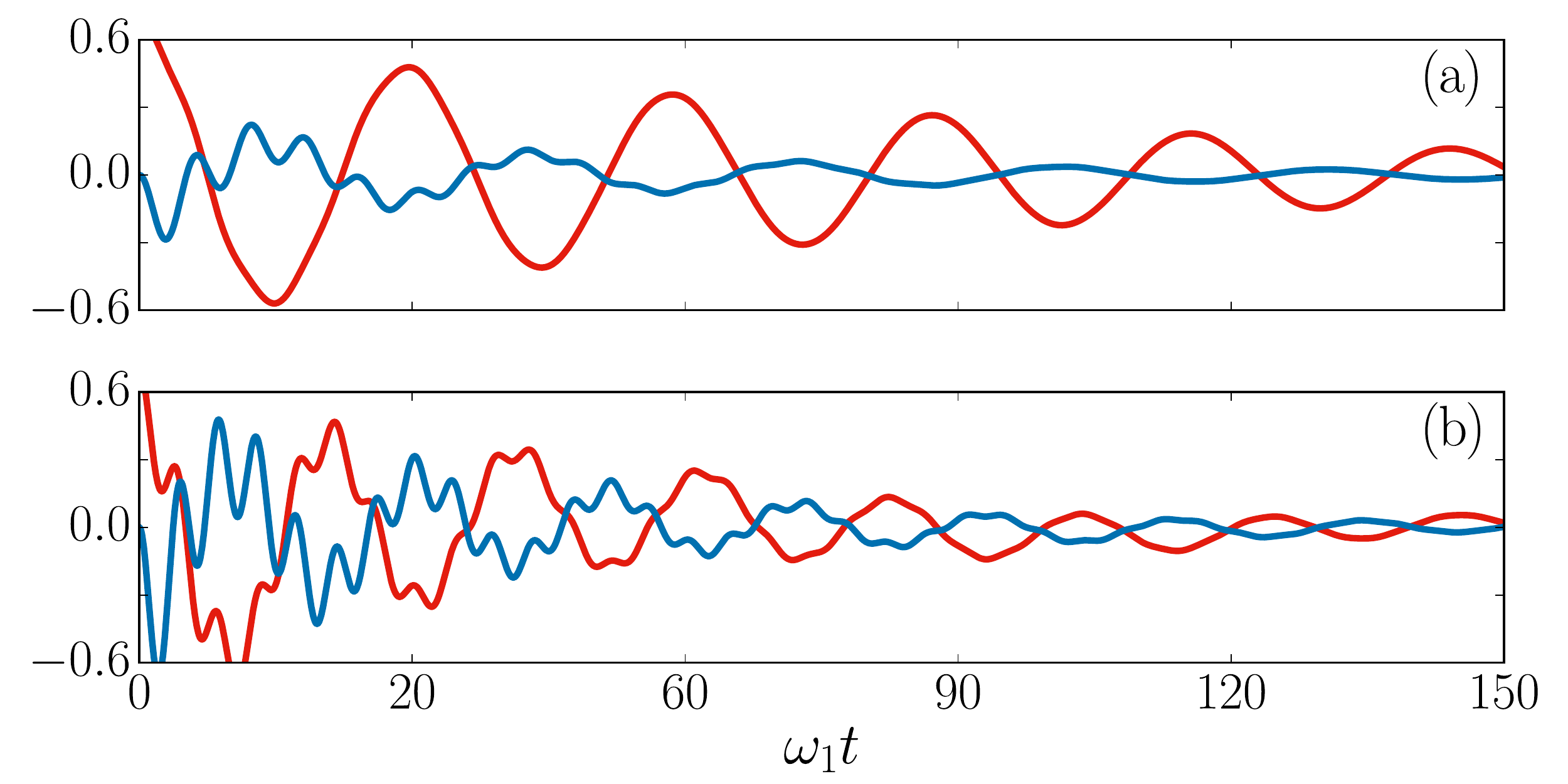}
 \caption{\label{figureS5} Synchronized trajectories for spins $\langle \sigma_2^x \rangle$  (red) and  $\langle \sigma_3^x \rangle$  (blue), with initial condition $|\Psi(t=0)\rangle=|0\rangle/\sqrt{2}+(|\{1\}_2\rangle+\{1\}_4)/2$. In  (a) we fix $\lambda=0.1\omega_1$, $\delta=0.75\omega_1$ (SS II), while in (b) $\lambda=0.475\omega_1$, $\delta=0.85\omega_1$ (SS I). In both cases we have $N=4$ and $\gamma_j=0.05\omega_j$.}
\end{figure}

In Fig. \ref{figureS5} we show two examples of synchronized trajectories. In (a) we plot a case in region II, while in (b) a case in region I. We only show two spin's coherences, $\langle \sigma^x_1\rangle$ and $\langle \sigma^x_2\rangle$, for clarity, although all of them are synchronized. In both cases we can see that after a transient, the spins synchronize almost in anti-phase and at the slow frequency, corresponding to the eigenmode with smallest decay rate.

\subsection{Synchronization maps for various dissipation strengths}\label{sect4b}

\begin{figure}[H]
 \centering
 \includegraphics[width=0.95\columnwidth]{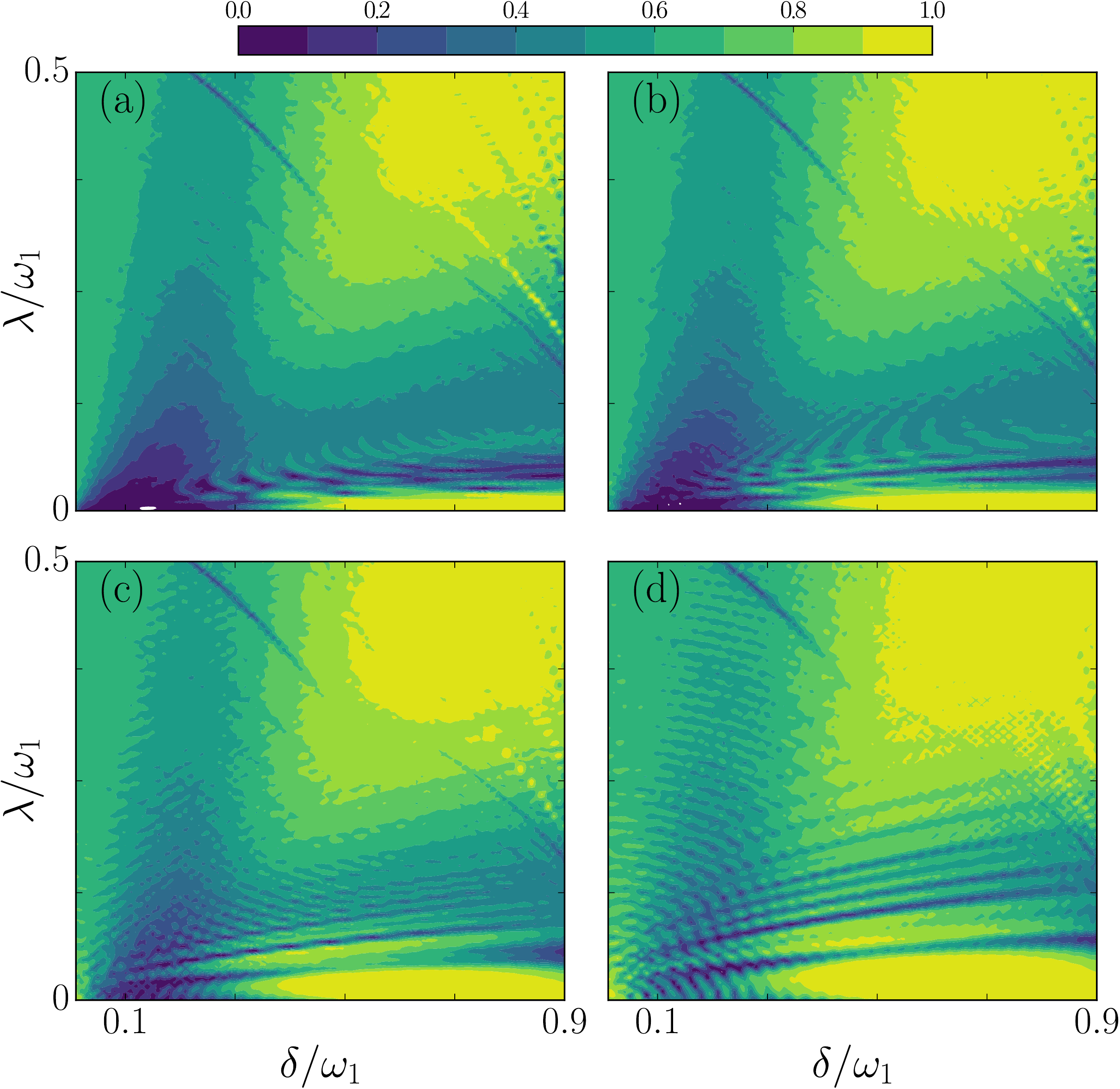}
 \caption{\label{figureS2} Map of SS among all spin pairs $\mathcal{C}_T(t)=\prod_{i<j}\mathcal{C}_{\langle \hat{\sigma}^x_i\rangle,\langle \hat{\sigma}^x_j\rangle}(t)$
at $\gamma_1 t=10$ and $\omega_1 \Delta t=80$, varying detuning and coupling strength, 
 and with $|\Psi(t=0)\rangle=|0\rangle/\sqrt{2}+(|\{1\}_2\rangle+|\{1\}_3\rangle)/2$. 
 Strong SS is found at the yellow (light color) regions I ({\it intra-band} synchronization) II ({\it inter-band} synchronization) ($\mathcal{C}_T\ge0.9$). For all figures $\omega_1=1$, while in (a) $\gamma_j/\omega_j=0.005$, in (b) $\gamma_j/\omega_j=0.01$, in (c) $\gamma_j/\omega_j=0.025$ and in (d) $\gamma_j/\omega_j=0.05$. (d) corresponds to Fig. 2(a) of the main text, and we have included it to ease comparison.}
\end{figure}

In this section we analyze the effects of varying $\gamma_j/\omega_j$ over the emergence of spontaneous synchronization (SS) for all the considered parameter region (Fig. 2(a) main text). In Fig. \ref{figureS2} we plot the synchronization map for increasing values of the ratio $\gamma_j/\omega_j$, (a)-(d), from $0.005$ to $0.05$ respectively. Comparing these plots, we observe that the main difference is the change in size of region II of synchronization (small $\lambda$ and large $\delta$), which diminishes with the ratio $\gamma_j/\omega_j$. Indeed, the value of $\lambda$ above which SS II is no longer found diminishes strongly, while the range of $\delta$ for which there is SS does not change significantly. Conversely, the other regions of the map do not change significantly when varying the dissipation strength. The decreasing size of region II is explained by recalling the mechanism behind SS in this region. As we explain in the main text, SS in region II emerges because the small difference between 
the eigenfrequencies ($\nu_l$'s) of the same band is blurred by the decay rates ($\Gamma_l$'s), resulting in an effective two body behavior (Fig. 3(d)  main text). When decreasing $\gamma_j/\omega_j$, the $\Gamma_l$'s become smaller relative to the $\nu_l$'s, and the dynamics of the system resolves better small frequency differences. This implies that the effective two-body behavior will be lost for smaller values of $\lambda$, hence hindering SS.

\subsection{Synchronization in larger chains}\label{sect4c}

\begin{figure}[H]
 \centering
 \includegraphics[width=0.95\columnwidth]{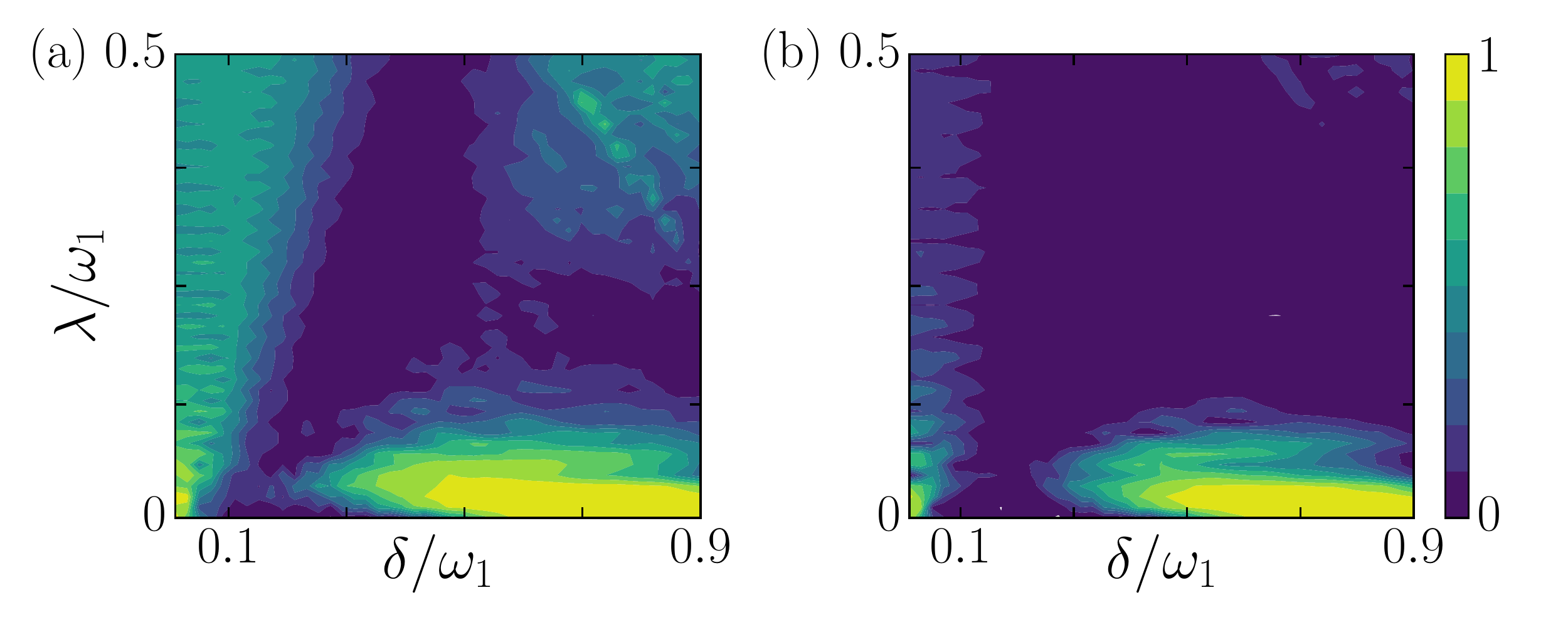}
 \caption{\label{figureS3}   Synchronization map
 as product of Pearson correlation between spin pairs coherence operators $\mathcal{C}_T(t)=\prod_{i<j}\mathcal{C}_{\langle \hat{\sigma}^x_i\rangle,\langle \hat{\sigma}^x_j\rangle}
 (t)$ at $\gamma_1 t=10$ and $\omega_1 \Delta t=80$, varying detuning and coupling strength, and with $|\Psi(t=0)\rangle=|0\rangle/\sqrt{2}+\sum_{j=1}^N|\{1\}_j\rangle/\sqrt{2N}$. 
 Strong SS is found at the yellow regions ($\mathcal{C}_T\ge0.9$). (a) For a chain of $N=6$ spins. (b) For a chain of $N=8$ spins. For both figures  $\gamma_j=0.05\omega_j$.}
\end{figure}

In this section we present the synchronization maps for larger chains of $N=6$ and $N=8$. Comparing figure \ref{figureS3} (a) and (b) with Figure 2(a) of the main text, 
we observe how synchronization in region I rapidly disappears as the size of the chain is increased. This is due to the fact that  it
depends on the difference between eigenvalues of the same band, which tends to vanish for increasing size $N$.
In contrast synchronization of region II is rather robust as it depends on the fixed 
gap $\delta$.  

\subsection{Two-time correlation functions in synchronized and unsynchronized regions}\label{sect4d}

\begin{figure}[H]
 \centering
 \includegraphics[width=0.95\columnwidth]{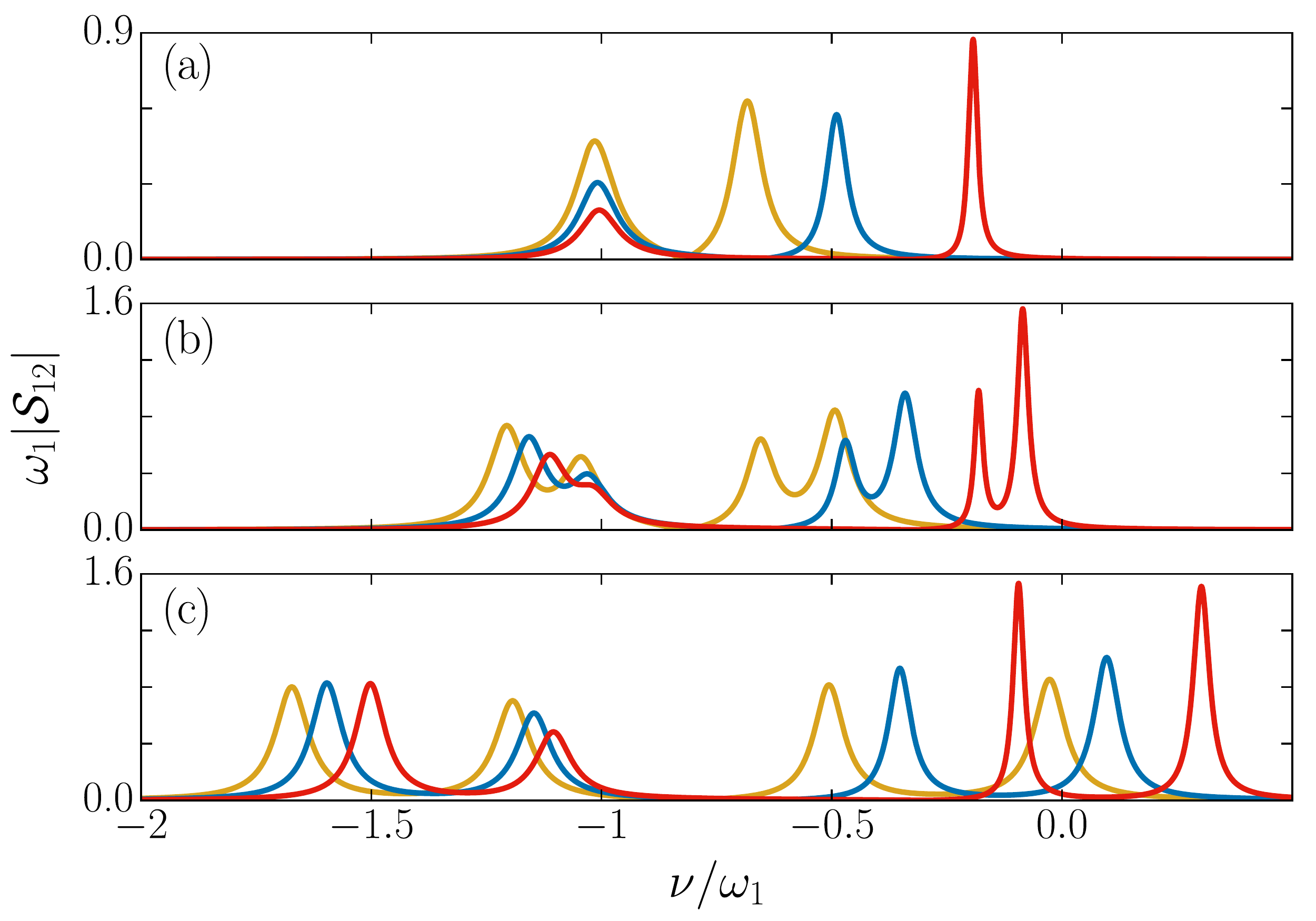}
 \caption{\label{figureS4} $\omega_1|\mathcal{S}_{12}(\nu/\omega_1)|$ for different couplings and detunings. In (a) $\lambda=0.05\omega_1$, in (b) $\lambda=0.2\omega_1$,  and in (c) $\lambda=0.5\omega_1$. Different line colors correspond to different detunings: gold $\delta=0.3\omega_1$, blue $\delta=0.5\omega_1$, red $\delta=0.8\omega_1$. In all cases $N=4$ and $\gamma_j=0.05\omega_j$.}
\end{figure}

In Fig. \ref{figureS4} we plot several examples of the two-time correlation 
given in (\ref{spectrum_eq}) for $j=1$ and $j'=2$. In (a) we plot examples for 
weak coupling ($\lambda=0.05\omega_1$), in (c) for strong coupling 
($\lambda=0.5\omega_1$) and in (b) for a coupling strength in which there is no 
synchronization ($\lambda=0.2\omega_1$). We plot in different colors three 
different detunings: $\delta=0.3\omega_1$ (gold), $\delta=0.5\omega_1$ (blue), 
$\delta=0.8\omega_1$ (red). In Fig. \ref{figureS4}a there are effectively only 
two peaks, and we can appreciate how when increasing the detuning one of the 
peaks becomes significantly thinner, indicating SS of region II. In contrast, if 
we increase the coupling to $\lambda=0.2\omega_1$, four different peaks emerge 
(Fig. \ref{figureS4}b). In this case there are two peaks which become thiner as 
detuning is increased, however both of them display a similar width, hindering 
SS. In contrast, for stronger coupling (Fig. \ref{figureS4}c), these two peaks 
display 
clearly different widths. This asymmetry in the decay rates of the eigenmodes is 
what enables synchronization in region I. 

We finally mention that this kind of two-time 
  correlations has also proven useful in the study of {\it stationary} synchronization, reported in Ref. \cite{S_SyncAtomicEnsemb}
  for two detuned and optically pumped clouds of atoms. Stationary SS arises for certain parameters and is characterized by the presence of 
  only one frequency, i.e. the presence of a single peak.   
Thus, transient SS response to fluctuations or to preparation in an out of equilibrium 
state are oscillations at multiple frequencies at first, but at just one 
frequency after a transient while in the pumped system  \cite{S_SyncAtomicEnsemb} 
oscillations occur at a single frequency. The illuminating quantity is the two-time 
correlation spectrum   as it provides a full characterization of the frequencies of the system. This quantifies not only stationary SS \cite{S_SyncAtomicEnsemb} (leading to a single peak)
but can also signal the presence of multiple dissipative time scales, crucial for the 
 presence of transient synchronization.

\begin{thebibliography}{99}
\bibitem{ordering} The eigenfrequencies (decay rates) given by (\ref{Kspec}), are  denoted by $\nu_l$ ($\Gamma_l$) with $l\in[1,N]$. The first $N/2$ elements correspond to the band '$-$' with the $k\to l$,  while the next $N/2$ to the band '$+$' with $k \to l-N/2$.
\end{thebibliography}
\end{document}